\documentstyle[12pt,epsf,epsfig]{article}
\def\be{\begin{equation}}
\def\ee{\end{equation}}
\def\ba{\begin{eqnarray}}
\def\ea{\end{eqnarray}}

\def\bq{\begin{quote}}
\def\eq{\end{quote}}

 at 10truept

\newcommand{\beq}{\begin{equation}}
\newcommand{\eeq}{\end{equation}}
\newcommand{\beqa}{\begin{eqnarray}}
\newcommand{\eeqa}{\end{eqnarray}}

\def\ltap{\ \raise.3ex\hbox{$<$\kern-.75em\lower1ex\hbox{$\sim$}}\ }
\def\gtap{\ \raise.3ex\hbox{$>$\kern-.75em\lower1ex\hbox{$\sim$}}\ }
\def\gl{\ \raise.5ex\hbox{$>$}\kern-.8em\lower.5ex\hbox{$<$}\ }
\def\roughly#1{\raise.3ex\hbox{$#1$\kern-.75em\lower1ex\hbox{$\sim$}}}

\def\d{{d}}
\newcommand\eqref[1]{(\ref{#1})}

\begin{document}

\thispagestyle{empty}
\begin{flushright}
March 2010
\end{flushright}
\vspace*{2cm}
\begin{center}
{\Large \bf McVittie's Legacy:  Black Holes in }\\
\vskip.3cm
{\Large \bf an Expanding Universe}\\
\vspace*{1.8cm} {\large  Nemanja Kaloper$^{\dagger ,}$\footnote{\tt
kaloper@physics.ucdavis.edu}, Matthew Kleban$^{\ddagger ,}$\footnote{\tt
mk161@nyu.edu} and Damien Martin$^{\dagger ,}$\footnote{\tt djmartin@ucdavis.edu}
}\\
\vspace{.5cm} {\em $^{\dagger}$Department of Physics, University
of California, Davis,
CA 95616}\\
\vspace{.5cm} {\em $^{\ddagger}$ 
CCPP, Department of Physics, New York University, 
New York, NY 10003}\\
\vspace{.15cm} \vspace{1cm} ABSTRACT
\end{center}

We prove that a class of solutions to Einstein's equations---originally discovered by G. C. McVittie in 1933---includes regular black holes embedded in Friedman-Robertson-Walker cosmologies.  If the cosmology is dominated at late times by a positive cosmological constant,  the metric is regular everywhere on and outside the black hole horizon and away from the big bang 
singularity, and the solutions asymptote in the future and near the horizon to the Schwarzschild-de Sitter geometry.  For solutions without a positive cosmological constant the would-be horizon is a weak null singularity.

\vfill \setcounter{page}{0} \setcounter{footnote}{0}
\newpage

\section{Introduction}

Finding solutions to Einstein's equations describing anything beyond the simplest and most symmetric configurations of matter or gravity is a hard and uncertain affair. When available they deserve our full attention, as they can provide us with new insights into the nature of gravity 
beyond the linearized regime or be of use in describing some objects of astrophysical relevance.  For both reasons any solution potentially describing a black hole embedded in an expanding universe is of considerable interest.

It is thus a bit surprising that a proper understanding of a class of  solutions found over 70 years ago by McVittie \cite{mcvittie} is still lacking. These solutions have many of the features one would expect of a black hole embedded in a Friedmann-Robertson-Walker (FRW) cosmology:  they are spherically symmetric, parametrized by a function $a(t)$ and a mass parameter $m$, reduce to FRW cosmology with scale factor $a(t)$ at large radius, and reduce to known black hole metrics or standard FRW cosmology in all the appropriate parametric limits.  These properties make them not only interesting in their own right as non-linear solutions, but potentially significant and physically relevant for describing real gravitating objects or holes in our universe.  On the other hand there is no accretion (the mass $m$ is constant)---an odd property for a physical black hole in a universe full of matter or radiation.

The most recent and thorough examination of the properties of the McVittie metric was attempted in a series of three papers by Nolan \cite{nolan}, which review past work on the metric and describe many of its features.  Because it is spherically symmetric and asymptotically FRW, the McVittie metric can be used to describe the external fields of finite size objects \cite{nolan,krasin,masses}, or exteriors of bubbles separating different regions of spacetime \cite{haha,bekutk}. In these applications the McVittie metric is replaced by a different geometry at small radius. 
But---as Nolan reviews at length---the literature on the McVittie solution proper is riddled with basic errors and contradictory statements, which has probably contributed significantly to its relative obscurity. 

Unfortunately Nolan himself erred on some points that are critical for the correct interpretation of the solution. Specifically, he argued (but didn't prove or derive) that the would-be  null black hole horizon of the McVittie metric is at {\em infinite} distance and therefore constitutes a null boundary rather than a horizon, and hence that the metric outside this surface is geodesically complete and cannot describe a black hole at all.  Much of the rest of his analysis relied on this assertion, which as we will demonstrate explicitly is {\em incorrect}.  The null surface is at a {\em finite} distance and therefore renders the standard form of the McVittie metric geodesically incomplete, a conclusion that entirely changes the interpretation  given in \cite{nolan} of the global structure of the spacetime and validates the black hole interpretation in at least some cases.

The cases for which the black hole interpretation is valid occur when the McVittie scale factor asymptotes to de Sitter space\footnote{If one takes a negative cosmological constant the cosmology must asymptote to an open FRW universe which originates from a big bang singularity and terminates in a big crunch singularity (see, e.g. \cite{kalinde} for a discussion). Similarly, in collapsing FRW universes with 
any value of the cosmological constant there would be
spacelike or null singularities in the future.  We will not consider these cases further.}
($a(t) \rightarrow e^{H_{0} t}$). The late-time exponential expansion turns out to dilute the FRW fluid(s) near the horizon rapidly enough that {\em all} curvature invariants reduce to their corresponding values for a pure de Sitter-Schwarzschild hole.  While we are unable to explicitly construct the analytic extension across the horizon, using the null ingoing geodesics we can show that the metric is nondegenerate and analytic near the horizon, which implies the existence of such an extension. In contrast, for generic power law scale factors ($a(t) \sim t^{p}$ at late times) the expansion is not fast enough to sufficiently dilute the stress-energy on the would-be horizon, rendering it weakly singular. We explicitly construct a higher derivative curvature invariant that diverges on the null surface in this case, implying the existence of infinite tidal forces there (and potentially large quantum corrections arising from higher curvature invariants in the effective action).

To conclude this introduction we list the salient features of a spatially flat McVittie metric  characterized by the scale factor $a(t)$ of the asymptotic FRW region and a mass $m$:

\begin{itemize}

\item for $m=0$, reduces to standard homogeneous and isotropic FRW cosmology with scale factor $a(t)$ and zero spatial curvature;
\item for $H(t) \equiv \partial_{t} \ln a(t) = $ constant, reduces to a Schwarzschild or de Sitter-Schwarzschild black hole of mass $m$ and Hubble constant $H$;
\item  has a spacelike and inhomogeneous big bang singularity in the past (for generic expanding $a(t)$);
\renewcommand{\labelitemi}{$\diamond$}
\item has a null or spacelike FRW future infinity at large radial distance $r$ and late time $t$;
\item has a surface $r=r_{-}, t=\infty$ which is null and  at {\em finite}  affine distance along ingoing null geodesics from any point in the bulk;
\item this null surface is a soft, null naked singularity in an FRW spacetime if the FRW Hubble constant $H(t) = \dot a(t)/a(t)$ goes to zero at late times;
\item the addition of any positive cosmological constant (so that $\lim_{t \rightarrow\infty} H(t) \equiv H_{0}  >0$) eliminates this singularity;
\item all curvature invariants on the null surface in this case are exactly equal to their values on the horizon of a Schwarzschild-de Sitter black hole of mass $m$ and Hubble constant $H_{0}>0$;
\item therefore, at least in the case  $H_{0}>0$ the McVittie metric describes a regular (on the horizon) black hole embedded in an FRW spacetime.
\end{itemize}

The items marked $\diamond$ are new results of the work reported in this paper, correcting existing claims in the literature.

\section{McVittie Retrospective}

We begin with a brief review of the McVittie solutions.  In their simplest form they have zero spatial curvature in the asymptotically FRW region, but  can be easily generalized to include non-zero positive or negative spatial curvature \cite{mcvittie} or electric charge
\cite{vaidya,kastra}. We do not expect that the spatial curvature of the FRW geometry to significantly alter the behavior of the metric near a mass source as long as the gravitational radius of the mass $m$, or the spatial extent of the region occupied by it, whichever is larger, is smaller than the radius of curvature.\footnote{In the first of references \cite{nolan}, Nolan concludes that open McVittie solutions are {\em not} good solutions describing masses in expanding FRW universes. We suspect that this inference arises from excessively strong assumptions.} Since this is presumably the case for astrophysical masses, we specialize to the case of zero spatial curvature. 

The solution is given by the metric
\be ds^2 = - \Bigl(\frac{1-\mu}{1+\mu}\Bigr)^2 dt^2 + (1+\mu)^4
a^2(t) d\vec x^2 \, , \label{mcvitt} \ee
where $a(t)$ is the asymptotic cosmological scale factor, 
\be \mu = \frac{m}{2a(t) |\vec x|} \, , \label{mu}
\ee
$m/G_{N}$ is the mass of the source, and using spatial translations we have chosen $\vec x = 0$ as the center of spherical symmetry. This is an exact solution of the field equations of Einstein's General Relativity
for an arbitrary mass $m$ provided that $a(t)$ solves the
Friedmann equation 
\be 3H(t)^2 = 8 \pi G_N \rho(t) \, , \label{friedman} \ee
where $\rho$ is the energy density $T^0{}_0$ and $H = \dot a/a$.  Surprisingly the energy density is constant along slices of constant $t$.  It scales with the cosmic scale factor and controls the overall expansion rate of the universe exactly as in a standard FRW geometry with scale factor $a$ and Hubble parameter $H$.  In fact we can linearize the McVittie solution by considering the limit $\mu \ll 1$, in which case the metric (\ref{mcvitt}) reduces to a perturbed FRW universe in Newtonian gauge, $ds^2 = -(1-2 \frac{m}{a|\vec x|}) dt^2 + a^2 (1+2 \frac{m}{a|\vec x|} ) d\vec x^2$.  Starting from this we could use linear superposition to build an arbitrary `multicentered' solution. The result is just a general perturbed FRW cosmology.

The pressure on fixed $t$-slices of the geometry (\ref{mcvitt}) is {\it not} homogeneous.
After a slight rewriting of the formula in e.g. \cite{sakaihaines},
\be
p = \frac{1}{8\pi G_N} \Bigl( - 3 H^2  -2 \frac{1+\mu}{1-\mu} \dot H  \Bigr) 
\label{pressure}
\ee
and so has two contributions: a homogeneous term $\propto H^2$, and an inhomogeneous term 
$\propto \dot H$. To understand the role of the second term, note that (except in the case of pure cosmological constant $\dot H=0$) one expects the mass to break the homogeneity of the stress-energy on spatial slices.  It should pull matter in from the FRW fluid around it, making the energy density inhomogeneous.  This does not happen in the McVittie solution: the energy density (\ref{friedman}) is a function of the cosmic time alone.\footnote{We call this time coordinate the `cosmic time' since it reduces to the usual comoving FRW time when the mass source is absent, and asymptotes to it far away from the source when the mass does not vanish.}  Therefore something must cancel the gravitational attraction of the mass, and that non-gravitational balancing force is provided by  the gradient of the pressure (\ref{pressure}).  

The McVittie solution has a curvature singularity at $\mu = 1$ (which can be seen by evaluating e.g. the Ricci scalar $R = 12H^{2}+6 \frac{1+\mu}{1-\mu} \dot H$) where the pressure goes to infinity.  As we will see this singularity is spacelike, extends all the way to spatial infinity, and should be viewed as a cosmological big bang singularity. 

One may wonder what the status of this solution is among candidate metrics describing black holes embedded in expanding FRW universes.
Nolan \cite{nolan} offers a theorem proving that the McVittie solution is the unique solution describing the field of a spherically symmetric mass in a spatially flat asymptotically FRW cosmology. However, his theorem relies on rather strong assumptions,
among which are that {\em i)} the mass is constant, and that {\em ii)} there exists a gauge in which energy density is spatially homogeneous despite the gravitational field of the central mass.  Evidently these conditions can only be attained with the help of the non-gravitational forces encoded by the inhomogeneous pressure (\ref{pressure}). 

These assumptions are by no means sacrosanct---on the contrary, they are very restrictive and not very well-motivated physically. Astrophysical black holes generally accrete matter at some non-zero rate.  Even solutions with no accretion should exist without the necessity of introducing non-gravitational forces, for example by assigning the matter an initial velocity distribution such that it flows radially away from the black hole at or above escape velocity.  In fact there exist special solutions in the ``Swiss cheese'' class \cite{schuck} where a region of flat FRW dust-dominated cosmology is excised and replaced by a void containing a black hole with a mass equal to that of the would-be mass of the excised region \cite{sussman}. One can imagine generalizing such solutions further by adding spherical dust shells to the void, each with a nonzero peculiar velocity so that {\it e.g.} it is marginally bound to the black hole and hence never falls in.  Presumably one could take a continuum limit of many such shells, finding a solution where the central mass remains constant, with zero pressure, but with the energy density a function of radius and time. 

Thus, the McVittie solution should be thought of as a special case of a larger class of geometries describing masses in FRW:  McVittie is the special class where the mass parameter is a constant {\it and} the energy density is homogeneous, and its inhomogeneous pressure is the necessary and sufficient price one pays for these features. 
 
The initial Big-Bang singularity is absent when $\dot H =0$, and in fact the geometries (\ref{mcvitt}) reduce to the Schwarzschild or Schwarzschild-de Sitter solutions (for $H=0$ and $H \ne 0$ respectively). The hypersurface $\mu=1$ is perfectly regular in those cases, being the event horizon in the Schwarzschild case, and a spacelike surface inside the event horizon in the Schwarzschild-de Sitter geometry. Their black hole singularities remain censored by the event horizon, and cannot be seen by exterior observes. 

\section{Coordinate Covers}

In the case $a(t)=1$ the McVittie solution reduces to a black hole in flat space;  setting $a=1$ in the metric (\ref{mcvitt}) gives the Schwarzschild solution in isotropic coordinates.  These coordinates have the unfortunate feature that the coordinate $|\vec x|$ covers the exterior of the back hole twice:  $m/2 < | \vec x| < \infty$ covers the same region---the exterior of the hole---as $0 < | \vec x | < m/2$.  

We will use another coordinate choice which more closely imitates the familiar (static) form of the Schwarzschild or Schwarzschild-de Sitter metric.  The new radial coordinate is defined by \cite{nolan}
\be
\vec r = (1+\mu)^2 a(t) \vec x \, , 
\label{eq:sds:static_time}
\ee
where 
$r = |\vec r|$ turns out to be the `spherical area' coordinate. 
This is `one-half' of the coordinate transformation which goes from cosmological to static patch coordinates in de Sitter space. The analogue of the static patch time is not so easy to find, because in general the true time dependence of the McVittie metric makes the integration harder.  One can show that a time coordinate analogous to static time does exist, but we will not need it for our analysis and will use $t$ as a time variable. 

The equation 
relating $a(t) |\vec x|$ to $r$ is quadratic, so the coordinate transformation 
(\ref{eq:sds:static_time}) actually defines two separate branches:
\be
a(t) |\vec x| = \frac{m}{2} \left(\frac{r}{m} - 1 \pm \sqrt{\Bigl(\frac{r}{m} -1\Bigr)^2 -1} ~\right)^{-1} \, .
\ee
To check which of these is physically relevant we note that in the limit $r \rightarrow \infty$, the product 
$a(t) |\vec x| \rightarrow 0$ on the ``+'' branch. This branch covers the region of space between $a(t) |\vec x| = 0$, where the curvature blows up at $a(t) = 0$ for any finite $\vec x$,  and $a(t) |\vec x| = m/2$, which by (\ref{mu}) corresponds to $\mu = 1$ and is another spacelike curvature singularity, the McVittie big bang.  Since this patch terminates on  spacelike curvature singularities both in the past and in the future, we will ignore it hereafter and specialize to the ``-'' branch, where $r \rightarrow \infty$ implies $|\vec x| \rightarrow \infty$ at any finite $a(t)$, so that the geometry asymptotes at large $r$ to an FRW region. 

We are most interested in the solutions which contain both perfect fluid (or fluids) and positive cosmological constant components. If the fluid obeys the null energy condition, these solutions will be dominated by the cosmological constant at late times. One might expect based on the maximally extended Schwarschild-de Sitter geometry that this class of McVittie solutions has a maximal extension given by   the periodic repetition of the fundamental block we will describe (for details, see e.g. \cite{gibbonshawking}), but we will not pursue this further here.

\begin{figure}[hbt]
\begin{center}
\includegraphics[width=.6\textwidth]{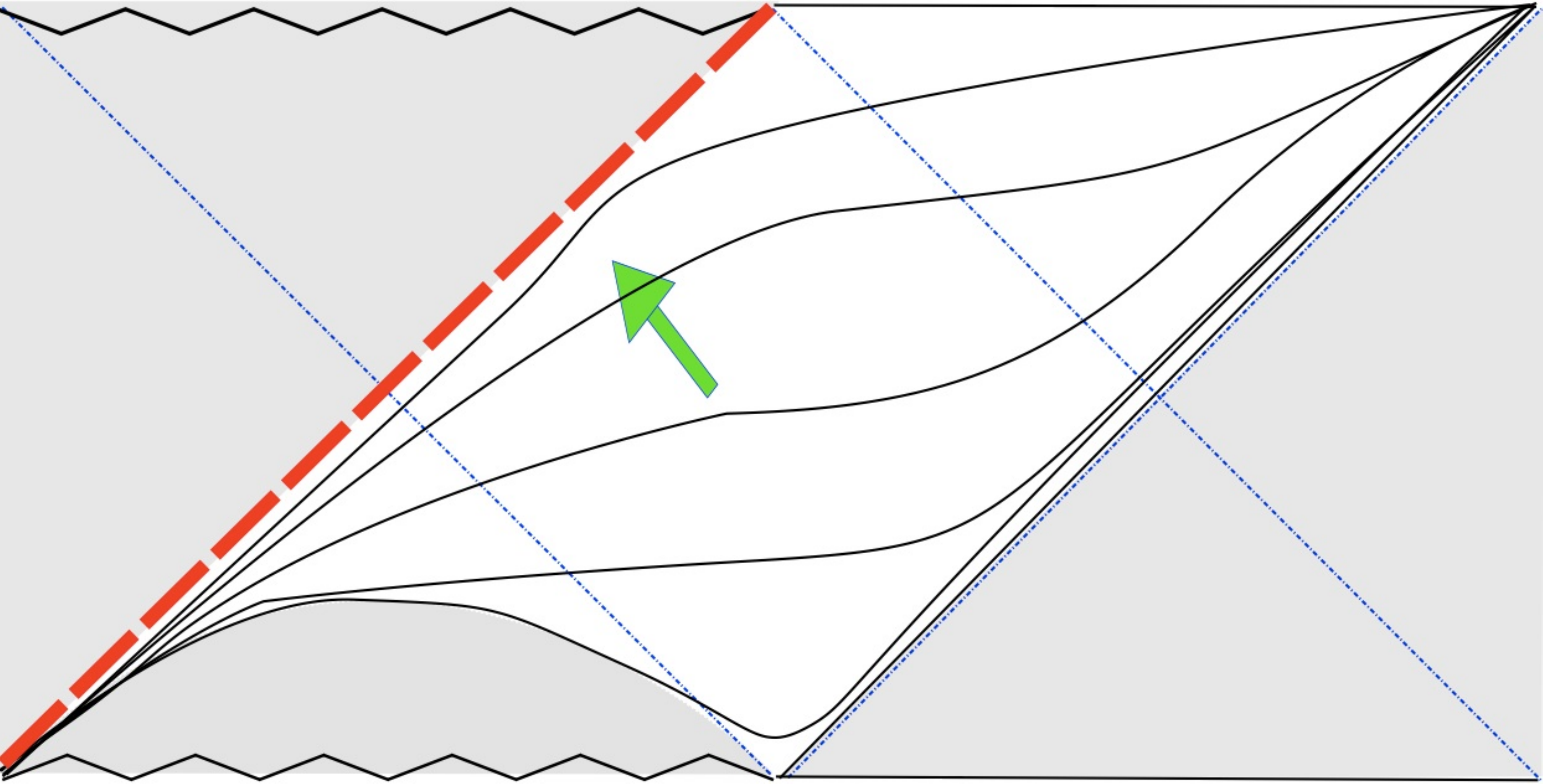}
\caption{The section of the Schwarzschild-de Sitter conformal diagram, which is covered by our coordinates $r, t$ when $H = H_0 = {\rm const}.$ Black solid lines are the surfaces of constant $t$, and the green arrow represents an ingoing observer, which crosses these surfaces en route to the bold red dashed line $r=r_-, t = \infty$, which is the null apparent horizon. This means that the coordinate $t$ is a good time coordinate for ingoing observers.}
\label{fig:confds}
\end{center}
\end{figure}

If we swap  $\vec x$ for $\vec r$ defined in the first of Eq.  (\ref{eq:sds:static_time}), the McVittie metric (\ref{mcvitt}) becomes 
\begin{equation}
\d s^2 = -f \d t^2 - \frac{2Hr}{\sqrt{1-2m/r}}\d r\,\d t + \frac{\d r^2}{1-2m/r}  + r^2 \d\Omega_2 \, ,
\label{cosmetric}
\end{equation}
where $f = 1- 2m/r - H(t)^2 r^2$.
When $H = {\rm const}$, this is the Schwarzsczhild-de Sitter metric in coordinates which are analogous to outgoing Eddington-Finkelstein coordinates for a flat space Schwarzschild black hole (see Figure \ref{fig:confds}). 

\section{Causal Structure}

We will now proceed to use the coordinates (\ref{cosmetric}) to analyze the causal structure of the McVittie solution, focusing on the case where $\rho$ consists of one or more perfect fluid components satisfying the null energy condition plus a positive cosmological constant (so that $\lim_{t\rightarrow 0}H = \infty$, and $\lim_{t \rightarrow\infty}H=H_{0}\geq 0$).   The metric function $f= 1- 2m/r - H(t)^2 r^2$ and its positive roots will play important roles in what follows.

There are five important surfaces that delineate the causal structure. 
 
 \begin{itemize}
 \item A surface at $r=2 m, t={\rm finite}$.  This is the curvature singularity at $\mu=1$ mentioned earlier ({\it c.f. e.g.} the Ricci scalar $R=12 H^{2}+6 \dot H /\sqrt{1-2m/r}$).   The metric (\ref{cosmetric}) shows immediately that it is a spacelike 3-surface:  since $r$ is fixed to $2m$, we have $dr=0$ and $g_{tt}=-f = 4m^2 H^2(t)>0$, and the sphere has finite radius $r=2m$.  This surface lies in the causal past of all spacetime points in the  patch of the metric we are studying, so we refer to it as the big bang.
  
\item A null apparent horizon at $r=r_{-}, t=\infty$, where $r_{-}$ is the smaller positive root of $f(t=\infty)=1- 2m/r - H_{0}^2 r^2=0$.  In the case $H_{0}>0$ we will demonstrate that this is a regular black hole event horizon. In the case $H_{0}=0$ it is a null singularity.  In both cases it is a finite distance from points in the interior.

\item A null surface ending at the point $r=r_{+}, t=\infty$, where $r_{+}$ is the larger positive root of $f(t=\infty)=0$.  For $H_{0}>0$ this is a cosmological event horizon. For $H_{0}=0$ it moves off to $r=\infty, t=\infty$ and becomes a null FRW infinity.

\item A spacelike apparent horizon at the smaller positive root $f=0$ with $t$ finite. It is connected to $\ldots$ 

\item $\ldots$ an apparent horizon at the larger positive root of  $f=0$ with $t$ finite. The two branches of apparent horizon at finite time $t$ link up at a bifurcation point and delimit a normal region of spacetime from an antitrapped region. In asymptotically de Sitter space, the outer branch of the apparent horizon asymptotes to the null surface through $r=r_+, t \rightarrow \infty$, whereas in asymptotically decelerating FRW cosmologies it ends at timelike infinity \cite{hks}. While the coordinate $t$ is everywhere timelike, the coordinate $r$ is spacelike only inside the normal region, enclosed by the apparent horizons.

 \end{itemize}

\begin{figure}[hbt]
\begin{center}
\includegraphics[width=.75\textwidth]{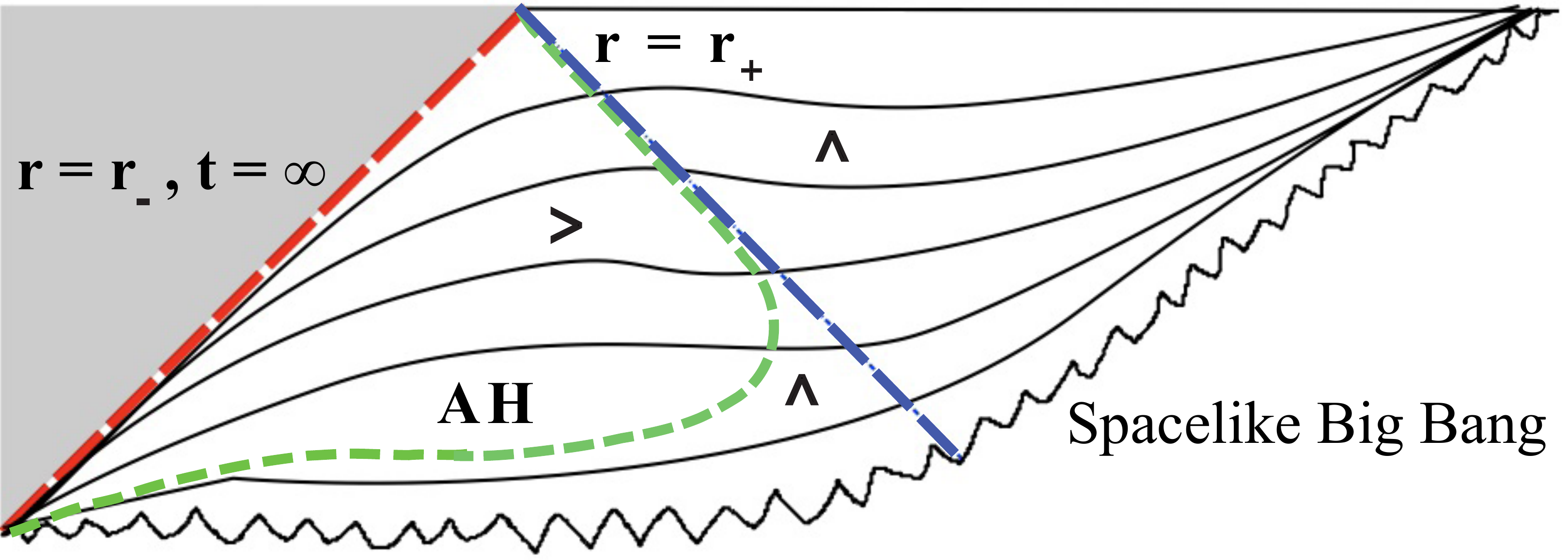}
\caption{The conformal diagram of the McVittie solutions which asymptote the Schwarzscschild-de Sitter geometry in the future. The red dashed line is the surface $r=r_-, t = \infty$, and the blue dashed line is the cosmological horizon through $r=r_+$. The broken curve at the bottom is the inhomogeneous Big Bang at $r=2m$ and finite times $t$, and the thin green dashed line is the union of the two branches of apparent horizons for finite $t$. Geodesic expansion is indicated by the now-standard `Bousso wedges'.}
\label{sdsmcvitt}
\end{center}
\end{figure}

\noindent The casual structure of the McVittie geometry is shown in Figure \ref{sdsmcvitt}.

\subsection{McVittie's Apparent Horizons}
\label{sec:apparent_horizon}

A feature of the geometry often helpful in understanding its causal structure is the apparent horizon, a surface where at least one congruence of null geodesics changes its focusing properties (see, e.g \cite{nolan, Hayward}). As it crosses the apparent horizon this family of geodesics flips from converging to diverging (or vice versa). So to find the apparent horizon, we seek the locus of a vanishing geodesic expansion. For any observer, this surface is the light front on which the area is extremal. In spherically symmetric spacetimes, the light fronts are spherical, and so it suffices to look for the extrema of the spherical area around the observer along radial null geodesics. 
These satisfy the null constraint $ds^2 = 0$ in (\ref{cosmetric}), which yields
\begin{equation}
\frac{\d t}{\d r} = \frac{\pm 1}{\sqrt{1-2m/r}(\sqrt{1-2m/r} \pm Hr)}\label{eq:radial_geodesics}
\end{equation}
where the upper (lower) signs apply for the outgoing (ingoing) rays.

The spherical area is just $A = 4\pi r^2$, so its variation (\ref{eq:radial_geodesics})  is 
$\delta A = 8\pi r \delta r = 16\pi r \frac{dt}{d\lambda} \frac{dr}{dt} \delta \lambda$, where $\lambda$ is the affine parameter along the null congruence, and $\frac{dr}{dt}$ is the inverse of (\ref{eq:radial_geodesics}). Since we are interested in {\it all} the apparent horizons, for arbitrarily oriented null geodesics (i.e. both in- and out-going), this equation shows that to find it we should find the composite hypersurface for which $A'=0$ either along ingoing or outgoing geodesics. Thus, the full apparent horizon is defined as the hypersurface on which 
\be
\left(\frac{dr}{dt} \right)_+ \, \left(\frac{dr}{dt}\right)_- = 0 \, ,
\ee
which yields 
\be
f = 1 - \frac{2m}{r} - H(t)^2 r^2 = 0 \, . 
\label{metfunct}
\ee

To find out how each individual family of geodesics behaves on and beyond the apparent horizon we need to be a bit more formal \cite{Hayward}. Define the geodesic expansion by considering the Lie derivatives of the area two-form perpendicular to the geodesic congurence. The two geodesic congruences are given by the flows of the vector fields $\partial_t +\left( \frac{dr}{dt}\right)_\pm \partial_r$, the area form for spherically symmetric wavefronts is just $r^2 \sin \theta d\theta d\phi$, and so the Lie derivatives are $\mathcal{L}_\pm \omega = \theta_\pm \omega$, where the geodesic expansions $\theta_\pm$ are
\be
\theta_\pm = \pm \frac{2}{r}\sqrt{1-\frac{2m}{r}}\left(\sqrt{1-\frac{2m}{r}} \pm Hr\right)\label{gexps}\ee
Where $\theta_\pm$'s are positive, the corresponding family of geodesics are diverging, and conversely, where they are negative the geodesics are converging.
The spacetime is naturally divided between the regions where both expansions are positive (anti-trapped) or negative (trapped), or they have different sign (normal). The question is, which of these regions may be accessed by observes who, at some point, can communicate with each other, and not have to face the doom of singularities separating them. 

In the limit of the Schwarzschild-de Sitter geometry, $H = {\rm const}$, the apparent horizon coincides with the event horizon (as is clear from (\ref{metfunct})). For a general McVittie solution the apparent horizon is given by the roots of (\ref{metfunct}), which express $r$ as a function of $t$ through the time-dependence of $H$. Evidently the area of the apparent horizon at any given time $t$ coincides with the Schwarzschild-de Sitter horizon area for a spacetime with the curvature given by the Hubble length {\it at that $t$}. As time goes on the area of the apparent horizon changes, in such a way so that the smaller root of (\ref{metfunct}) moves {\it inward} as $H$ decreases with time towards its asymptotic de Sitter value, while the larger root moves {\it outward}. 

We see this quantitatively from taking the time derivative of 
(\ref{metfunct}) and evaluating the rate of change of the radial extent of the apparent horizon. It is given by
\begin{equation}
\left.\frac{\d r}{\d t}\right\vert_{\mbox{horizon}} = \frac{2H\dot{H} r^3}{(1- \sqrt{3}H r)(1+\sqrt{3}Hr)} \, , 
\label{eq:horizon_evolution}
\end{equation}
where we have used (\ref{metfunct}) to eliminate terms $\propto m/r$. Now, $r = \frac{1}{\sqrt{3}H}$ is the value of the horizon radius for the extremal Schwarzschild-de Sitter solution. So we see that at small $t$ when $H > \frac{1}{3\sqrt{3} m}$ the apparent horizon does not exist. As $t$ increases its two spacelike branches bifurcate and, as noted above, move inward (to smaller $r$) and outward respectively. The inner branch evolves along spacelike directions, since the interval between its subsequent locations is
\begin{equation}
\d s^2 = (\d t)^2 \left(\frac{\d r}{\d t}\right)\left[-2 + \frac{1}{H^2r^2}\frac{\d r}{\d t}\right], \quad\quad \mbox{using }1-\frac{2m}{r} = H^2 r^2 \mbox{ on A.H.} \, ,
\end{equation}
and noting, from (\ref{eq:horizon_evolution}), that $\frac{d r}{d t} < 0$, such that $ds^2 >0$. The outer branch eventually becomes timelike, asymptoting to null in the distant future. 

So far we have investigated the solutions $r(t)$ of $f=0$ with $t$ finite.  But in the limit $t \rightarrow \infty$ at least one of the two solutions 
remains finite. Let $r_{\pm}$ be the two positive roots of
\be
f(t=\infty)=1 - \frac{2m}{r} - H_0^2 r^2,
\label{sdshor}
\ee
where $H_0$ is the asymptotic future value of the McVittie Hubble parameter.  
The solution $r=r_{-}, t=\infty$ is evidently a null apparent horizon, since in this limit $f=0$, $\dot H = 0$, and the norm of the normal vector to the surface vanishes.  The properties of this null surface are key to our analysis.  The pressure singularity in (\ref{pressure}) may be absent there, since $\dot H = 0$. To check, we must determine the rate at which $H$ asymptotes to its constant future value along infalling geodesics, and at the same time settle the question of whether ingoing null geodesics reach it at finite affine parameter.  We will do both in the next section.

The larger root $r=r_+, t=\infty$ is a point on the cosmological event horizon, defined by the past-directed, outgoing radial null geodesics that end at it, when $H_0 \ne 0$. Near it the geometry is controlled by the positive cosmological constant, and future-directed outgoing null geodesics are well behaved, crossing the horizon at finite values of the cosmic time $t$. Constant $t$ surfaces also cross it (since they are spacelike). In the limit $H_0 = 0$, this horizon goes to infinity and becomes FRW null infinity. Note, that the finite-$t$ branch of the apparent horizon at finite times---given by the larger root of (\ref{metfunct})---always remains {\em inside} this cosmological Schwarzschild-de Sitter horizon, asymptoting to it at late times. Thus, the apparent horizon becomes first timelike and then eventually null, as noted above and sketched in Figure \ref{sdsmcvitt}.

To complete our description of the the causal boundaries of the McVittie solutions, we need to see what happens in the past. In the region covered by our coordinates the cosmologically expanding McVittie spacetime (\ref{mcvitt}) starts from a past spacelike Big Bang singularity at $r=2m$ and approaches the de Sitter limit at $t\rightarrow \infty$ if $H \rightarrow {\rm const} \ne 0$, or some power law FRW cosmology if $H \rightarrow 0$. As already noted, the $r=2m, t=$ finite singularity is spacelike.  It appears to the future of the naive FRW singularity at $a=0$ (where  $H$ diverges).  

One way to see this is to recall that $r=2m$ corresponds to $a(t) |\vec x| = m/2$ in the coordinates (\ref{mcvitt}), which yields $t \sim 1/|\vec x|^{1/s}$ for the singular surface at early times (where $s>0$ depends on the matter equation of state). Thus the singular surface at $r=2 m$ approaches the surface $a(t) = 0$ at spacelike infinity from {\it above}. Since in the de Sitter-energy fluid geometry the constant $t$ surfaces asymptote towards the central observer's particle horizon as $\vec x \rightarrow \infty$, the $\mu=1$ surface bends up in a similar fashion far away (see, e.g. \cite{hks}). Closer in, this surface also gradually bends up (relative to $a(t)=0$ surface), while remaining both spacelike and below the apparent horizon at all finite times, eventually crossing the null surface $r = r_-, t = \infty$. 

So we see that the McVittie solution (\ref{mcvitt}) with $H_{0}>0$ describes the region bounded by the null surface $r=r_-, t = \infty$, the big bang singularity $r=2 m$, and the asymptotic cosmological region. We can map it onto the Schwarzschild-de Sitter conformal diagram as illustrated in Fig. (\ref{sdsmcvitt}), where the jagged line is the big bang. Manifestly the spacelike big bang and the cosmological sector in the upper right are a part of the spacetime, since the null geodesics (\ref{eq:radial_geodesics}) get to the cosmological horizon $r=r_{+}$ in finite affine parameter. 

In the next section we will concentrate on the null surface $r=r_-, t = \infty$, first showing that it can be accessed in finite affine parameter by ingoing null geodesics, and determine whether it is traversable or singular.

\section{$r=r_-, t = \infty$: a Surface Too Far?}

In a typical picture of black hole formation from collapsing matter the apparent horizon is always found to  be behind the event horizon, when matter satisfies the null energy condition. In contrast, in the McVittie geometries the apparent horizon extends outside the would-be event horizon.
This might tempt one to argue that the expanding McVittie solution cannot have any black hole event horizons, since the future-oriented ingoing null rays seem to cross the inner branch of the apparent horizon. Led by this, Nolan \cite{nolan} decided that the null surface $r = r_-, t = \infty$ is a null boundary of the spacetime, separated from any interior point by an infinite lapse of affine parameter along any causal geodesic, and that therefore the McVittie spacetimes are geodesically complete to future directed null geodesics.

Such situations are sometimes encountered as, for example, in extremal magnetically charged black hole spacetimes 
\cite{gibbmaeda,strominger} and some charged $3D$ black strings \cite{kalanderson}. However, they tend to be `unstable' in the sense that when another parameter is turned on, such as e.g. electric charge, the boundary moves inward and the spacetime becomes geodesically incomplete. Hence it would be surprising if the surface $r=r_-, t = \infty$ is a boundary in a general case, since this conclusion would have to hold for {\it any} external matter equation of state, in a completely generic fashion. Already in \cite{nolan} doubts were expressed about this, although apparently the author decided against them. 

The argument of \cite{nolan} can be invalidated quickly once one recalls that the ingoing geodesics that start at the Big Bang singularity and cross the apparent horizon in fact enter the {\em normal region} of spacetime. They are {\em converging}. The outgoing geodesics are not, but this does not say anything about the surface $r=r_-, t = \infty$. In fact the formulas for geodesic expansion (\ref{metfunct})
show that it is actually a null branch of the apparent horizon, and there a family of geodesics will change its convergence properties. It may happen that both are convergent beyond this surface, which would indicate a trapped region, and identify $r=r_-, t = \infty$ as the event horizon. Thus to investigate the nature of the surface one must check explicitly if null ingoing geodesics can access it in finite affine parameter.

In what follows we shall prove explicitly that they can, and hence that the surface $r=r_-, t = \infty$ is {\it not} a boundary and the spacetime is not geodesically complete to future ingoing null rays. We will not be able to explicitly solve the geodesic equations.\footnote{For example, in the case of single fluid solutions and vanishing cosmological constant, where $a = t^p$, the geodesic equations reduce to a single first order Abel's differential equation $ \frac{dY}{d\theta} = \Bigl(f + \frac{p}{2} \frac{df}{d\theta} \Bigr) Y^2  + \frac{pf^2}{2} Y^3$ with $f=\frac{\cosh^3(\theta)}{\sinh(\theta)}$, which, as far as we could ascertain, is not known how to integrate \cite{kamke}.} However an exact solution is not necessary because we can set a finite upper bound on the lapse of the affine parameter along ingoing null geodesics during the approach to  $r=r_-, t = \infty$.

To proceed we need the null radial geodesic equations. One is the first integral  (\ref{eq:radial_geodesics}). We get another by varying the geodesic ``Lagrangian'' 
${\cal L} = \frac{r'^2}{1-2m/r} - \frac{2Hrr' t'}{\sqrt{1-2m/r}} - ft'^2$, where prime denotes derivative with respect to the affine parameter. If we define two functions $F_\pm = \pm (\sqrt{1-2m/r} \mp Hr) t' + r'/\sqrt{1-2m/r}$, the Lagrangian factorizes as ${\cal L} = F_+ F_-$. This means that the geodesic equations will be of the form
\be
\Bigl( F_+  \partial_{q'} F_- + F_-  \partial_{q'} F_+ \Bigr)' = F_+ \partial_{q} F_- + F_-  \partial_{q} F_+ \, ,
\label{geodeqs}
\ee
where $q$ is either $t$ or $r$.  Specializing to ingoing null geodesics implies by (\ref{eq:radial_geodesics}) that we can set $F_-=0$ after the variation, reducing 
(\ref{geodeqs}) to only the terms involving derivatives of $F_-$. Choosing $q=t$ and using (\ref{eq:radial_geodesics}) gives
\ba
\frac{dt}{dr} &=& - \frac{1}{\sqrt{1-2m/r} (\sqrt{1-2m/r} - Hr)} \, , \nonumber \\
{r}'' &=& \frac{r \dot H {r'}^2}{\sqrt{1-2m/r}(\sqrt{1-2m/r} - Hr)^2} \, ,
\label{radeq}
\ea
where the $\propto t'$ terms in the second equation are eliminated using the first.

From Eqs. (\ref{radeq}) we can see by a simple argument that ingoing null geodesics reach the surface $r=r_-, t = \infty$ at a finite value of the affine parameter.  Combining Eqs. (\ref{friedman}) and (\ref{pressure}) gives
\be
{1 + \mu \over 1 - \mu} \dot H = -4 \pi G_{N} \left( \rho + p \right) \le 0 \, ,
\label{nullen}
\ee
where the inequality holds if the null energy condition is satisfied.  In this case the geodesic `acceleration' $r''$ in the region $\mu<1$ (the future of the big bang) is non-positive, and therefore an ingoing geodesic with initial geodesic `velocity' $r'_{0}<0$ can never turn around or decrease its speed. Starting at any initial distance $r > r_-$, it must reach $r_-$ after an affine `time' $\Delta \lambda \le (r-r_{-})/ |r'_{0}|$, which is finite. The first of Eqs. (\ref{radeq}) shows that in this limit $ t \rightarrow \infty$ as expected. This conclusion is completely independent of the asymptotic form of $H$ as long as the null energy condition is satisfied. {\em Therefore, all physical McVittie solutions (\ref{mcvitt})  are geodesically incomplete to null future oriented ingoing geodesics, in contradiction to the claim of \cite{nolan}.} 

This suffices to show that the McVittie spacetime (\ref{mcvitt}) is geodesically incomplete, but it can be refined in order to produce a more precise estimate of the rate at which the surface $r=r_-, t =\infty$ is approached. We will need this estimate in the analysis of when the null surface $r=r_-, t = \infty$ is traversable and is a horizon, and conversely when it is impassable and a singularity. We defer the details of this analysis to the Appendix \ref{app:infall}.

 \section{{\it In Medias Res}: the McVittie Event Horizon}

So, what is the null surface $r=r_-, t = \infty$?  We will show that in the case when 
$H \rightarrow H_{0}>0$,  it is a regular event horizon, and the McVittie solution really is a black hole, which asymptotes the Schwarzschild-de Sitter solution. To prove this, ideally we'd construct a family of ingoing null geodesics that cross the horizon. This family is parameterized by an integration constant $u$, and each line depends on the affine parameter $v = \lambda$. Given such a construction we could view the smooth parametric equations of the worldines
\be
r = r(u,v) \, , ~~~~~~~~~~~ t = t(u,v) \, ,
\label{nullparameters}
\ee
as a coordinate transformation, exchanging the original coordinates $r, t$ (with $t$ divergent and thus ill-defined on the horizon) with the new coordinates $u,v$ which are perfectly regular and well defined on the horizon and beyond. We cannot find such coordinates globally in closed form  due to the complexity of the nonlinear geodesic equations. However, we will now demonstrate that such coordinates do exist, and we can construct them at least locally in the near horizon region.  Further, once we have demonstrated that those coordinates exist, we will check that near the horizon they reduce to null coordinates which cover Schwarzschild-de Sitter, and use this to 
prove that all curvature invariants are finite at the surface, in contrast to McVittie solutions which have asymptotically vanishing Hubble parameter.\footnote{Checking that the curvature invariants are finite on a null surface is not enough to conclude that the surface is regular. An example is provided by wavy cosmic strings with null singularities which are invisible to any invariants \cite{kalrob}. However, our proof is stronger since we demonstrate the existence of a smooth family of ingoing null geodesics. The calculation of the curvature invariants then simply serves to show that there are no hairs on this surface, and that the geometry really asymptotes Schwarzschild-de Sitter there.} We will demonstrate this at the end of this section.

First off, a set of null coordinates which cover (\ref{cosmetric}) certainly exists, since by spherical symmetry the $r-t$ part of the metric is conformally flat. Next, the metric (\ref{cosmetric}) is clearly non-degenerate at $r = r_-, t = \infty$ when $H_0 > 0$, since at $f \rightarrow 0$ it reduces to
\be
ds^2 \rightarrow - 2 dr dt + \frac{dr^2}{H_0^2 r_-^2} + r_-^2 d\Omega_2 \, .
\label{ascosmetrics}
\ee
At this surface the coordinates $r, t$ are `almost' null, in the sense that the only term which spoils the null tilt of the $t$ coordinate is $dr^2/H_0^2 r_-^{2}$, and in addition there is the irritating fact that $t$ diverges. However,  the individual curvature components remain finite there.\footnote{Curvature components are easy to calculate using either (\ref{mcvitt}) or (\ref{cosmetric}). We give them for the $r, t$ coordinates and the metric (\ref{cosmetric}) in Appendix \ref{app:curvature}. One can verify that they are regular at $r=r_-, t = \infty$, and by the `staring' argument one can begin to suspect that any number of the covariant derivatives of this metric will remain regular...  at least as long as $H \rightarrow H_0 \ne 0$ on this surface.}

The unfortunate features of the coordinate $t$ can be corrected with a radially dependent time translation. Choose a new time coordinate
\be
d\tau = dt +  \frac{dr}{\sqrt{1-2m/r} (\sqrt{1-2m/r} - F(r) r)} \, ,
\label{newtime}
\ee
where $F(r) = H(t(r))$, and $t(r)$ is a one-parameter family of solutions of Eqs. (\ref{radeq}) describing ingoing null radial geodesics. This new time coordinate is well defined away from the extremal limit, and can be integrated directly, because the second term on the RHS of (\ref{newtime}) is a total derivative. Also a suitable coordinate exists in the extremal limit as well. 
This new time coordinate 
{\it does not} diverge near the surface $r=r_-, t = \infty$! Indeed, in terms of this coordinate the first of 
Eqs. (\ref{radeq}) reduces to
$
\frac{d\tau}{dr} = \frac{1}{\sqrt{1-2m/r}} \left( \frac{1}{\sqrt{1-2m/r} - F(r) r} - \frac{1}{\sqrt{1-2m/r} -H(t) r} \right)$, and so along an ingoing null geodesic, where we insert the function $t(r)$ in place of $t$,
\be
\Delta \tau = 0 \, .  \label{geodeqone}
\ee
Thus $\tau$ is a conserved quantity of a given null geodesic, and can only vary from one geodesic to another. Now, note that

\begin{itemize}
\item the transformation of coordinates (\ref{newtime}) is nonsingular everywhere away from $r=r_-$;
\item the ingoing null geodesics reach $r=r_-$ in finite affine geodesic parameter, as we have shown in the previous section and in more detail in the Appendix \ref{app:infall};
\item the family of ingoing null geodesics is smoothly parameterized by the affine parameter and an integration constant which indexes the geodesics.
\end{itemize}

This implies that the coordinate transformation (\ref{newtime}) remains nonsingular in the whole patch of the exterior geometry inclusive of the surface $r=r_-$. Because of this, the coordinate $\tau$, being orthogonal to the affine parameter $\lambda$ since it is conserved along any given geodesic, can only be a function of the integration constant in the solution (\ref{radeq}). This implies that $\tau$ is continuous, and so also finite, as claimed. Thus, the divergence of the time coordinate $t$ has been regulated away, and $\tau$ will be well-behaved -- smooth and bounded -- on the surface $r = r_-, t = \infty$. 

We can now rewrite (\ref{cosmetric}) in terms of the $\tau$ coordinate. After some straightforward algebra we obtain
\ba
ds^2 &=& - f d\tau^2 - \frac{2 d\tau dr}{\sqrt{1-2m/r}} \left(Hr - \frac{ f}{\sqrt{1-2m/r} - F(r) r}\right) \nonumber \\
&& +  \frac{dr^2}{1-2m/r} \left( 1 + \frac{2Hr}{\sqrt{1-2m/r} - F(r) r} - \frac{f}{(\sqrt{1-2m/r} - F(r) r)^2} \right) + r^{2 }d\Omega_2 \, , ~~~
\label{crossing}
\ea
where $f = 1-2m/r - H^2 r^2$ is still the same function $f$ we defined in (\ref{metfunct}), and $H$ now depends on both $\tau$ and $r$, through the integral $t(r,\tau)$ defined by (\ref{newtime}). When we take the limit $r=r_-, t = \infty$, thanks to the form of $F(r) = H(t(r))$ and the limiting form of the function $f$ when $r \rightarrow r_-$, this metric reduces simply to 
\be
ds^2 = 2 d\tau dr + r_{-}^{2}d \Omega_2 \, ,
\label{nulllimits}
\ee
which is non-degenerate.
The coordinates $\tau, r$ are regular and bounded at this surface. Further, $\tau$ is defined by a solution of differential equation (\ref{newtime}) which exists, and is analytic everywhere in the domain of interest. So the full metric of Eq. (\ref{cosmetric}) must also be analytic at $r=r_-, t=\infty$, when $H \rightarrow H_0 \ne 0$, and the surface $r=r_-, t=\infty$ must be a traversable horizon.  The
geometry (\ref{mcvitt})  really represents a black hole embedded in an expanding universe, which emerged from some singular, inhomogeneous initial data at the Big Bang singularity $\mu = 1$, so long as $H_{0}>0$.

What goes wrong with this argument when $H_0 = 0$, as for example when $a = t^p$?  The surface $r=r_{-}=2m$ is still null and still at finite affine distance.  However a glance at Eqs. (\ref{newtime}) and (\ref{crossing}) shows that the coordinate $\tau$ is {\em not} finite on the ``horizon'' in that case.  As we will see, that is not due to a bad choice of coordinates---the surface is actually singular.

To see this we begin by examining the covariant curvature components, Christoffel symbols, and inverse metric given in Appendix \ref{app:curvature}.
One can see immediately that the only singularities in these components occur when $r=2m$ and when $a(t)=0$.  Taking (ordinary) partial derivatives with respect to $t$ or $r$ can change the degree of divergence at these points, but cannot introduce divergences at any other points.  Since all possible curvature invariants are constructed by products of these functions and their ordinary partial derivatives, divergences can only occur on these two surfaces.

We have already seen that the surface $r=2m, t=$ finite is a big bang curvature singularity.  The surface $t=0$ (where $a$ vanishes) is to its past and not a part of our coordinate patch.  What about the surface $r=r_{-}, t=\infty$?  Since $r_{-}>2m$ when $H_{0}>0$, no curvature can diverge on the horizon, and therefore it is traversable and regular.

To evaluate the curvature on this surface in more detail we should take the limit as we approach it along an ingoing geodesic.  For the case $H_{0}>0$, the near horizon scaling is $t(r) \sim {\alpha \over H_{0}} \ln{r_{-} \over r-r_{-}}$, where $\alpha>0$, as we have shown in Appendix \ref{app:infall}.  Looking again at the ingredients of the curvature invariants, recalling that $\dot H \sim e^{-H_{0}t}$ and $r_{-}>2m$, one can see immediately that the only terms that survive in the limit are those proportional to $H$ and not to any of its time derivatives.  But $H=H_{0}$ on the horizon, and therefore  {\em all curvatures on the horizon are exactly equal to their values on the horizon of a Schwarzschild-de Sitter black hole of mass $m$ and Hubble constant $H_{0}$.}  

In fact, the observation that the inverse metric, Christoffel connection, the curvature tensors and all of their derivatives are regular on this surface in effect {\it guarantees} the existence of a nonsingular family of geodesics which cross this surface, since one can start in any arbitrarily small neighborhood of the surface and construct the geodesics using the power series method. This is why we were able to define the coordinate $\tau$ in (\ref{newtime}) and change the metric to (\ref{crossing}), which is regular across the horizon.

The case $H_{0}=0, a(t) \rightarrow t^{p}$ at large $t$ is qualitatively different.  Since $r_{-}=2m$, the above argument no longer guarantees that the curvatures are bounded on the would-be horizon.  On the contrary, we find that they are in general divergent.  The equation for an ingoing null geodesic in the near-horizon region (with $a(t) \sim t^{p}$ at large $t$) is $t \rightarrow p r/\sqrt{1-2m/r}$. To determine this scaling, susbstitute $a \sim t^p, H = p/t$ in the first of Eq. (\ref{radeq}). This gives
\be
\frac{dt}{dr} = - \frac{1}{\sqrt{1-2m/r} \Bigl(\sqrt{1-2m/r} - pr/t\Bigr) } \, .
\label{radpower}
\ee
Because the null geodesics do reach the surface $r=r_-, t = \infty$ in finite affine parameter, $t$ must diverge to infinity as $r \rightarrow r_-=2m$ when $H_0 = 0$, as above. Indeed, if this were not true, and $t$ remained finite as $r \rightarrow 2m$, null future oriented ingoing radial geodesics would have 
all ended on the Big Bang singularity - a {\em past spacelike} singularity - which is clearly impossible. Further, the sign of $\frac{dt}{dr}$ cannot change as $t$ increases, because if it did it would imply that the future evolution would cease, with null geodesics turning back in time as $r \rightarrow 2m$. This would imply the existence of closed timelike curves and breakdown of causality in the regions of McVittie solution (\ref{mcvitt}) which are obviously fine, and so it cannot happen. Hence, we see that $1/t$ must follow $\frac{\sqrt{1-2m/r}}{pr}$ from {\em below}. Setting $\epsilon = \sqrt{1-2m/r} - pr/t \ge 0$, solving for $t$ 
and taking its derivative to the leading order in $\epsilon$, and comparing it to the formula
(\ref{radpower}), that translates to $\frac{dt}{dr} = - (\epsilon \sqrt{1-2m/r})^{-1}$, we can explicitly confirm this scaling, when we pick
\be
\epsilon = A \Bigl(1-2m/r\Bigr) + \ldots\, ,
\label{epsilon}
\ee
for $A$ a constant and ellipsis designating subleading terms. 

Using this scaling one sees that the Ricci scalar $R=12 H^{2}+6 \dot H /\sqrt{1-2m/r}$ is finite in the limit.  However one may suspect that higher curvature invariants will not be, since by taking a sufficient number of partial derivatives with respect to $r$ one should be able to obtain a divergence from the square root term.  To be certain there are no miraculous cancellations, we computed the invariant $\Xi \equiv (\nabla_{\mu} \nabla_{\nu} R_{\rho \sigma \delta \lambda})^{2} $ (see Appendix \ref{app:xi} for the result).  It contains terms such as $H^{4} (\dot H)^{2} (1-2m/r)^{-5}$ which diverge at the horizon along ingoing null geodesics, and there is no cancellation among the divergent terms.

An alert reader may by now be a little discomforted by the following observation.  On the one hand we are claiming that all curvatures are finite on the horizon when $H_{0}>0$, and have a finite limit when $H_{0} \rightarrow 0$ (since the horizon curvatures of a de Sitter-Schwarzschild black hole of mass $m$ simply reduce to those of a Schwarzschild black hole of the same mass in that limit, which are themselves manifestly finite).  On the other hand we claim that certain curvature invariants diverge when $H_{0}=0$.  How can this be? 

The resolution of this puzzle is that some curvature invariants, such as $\Xi$, do not monotonically decrease along ingoing null geodesics when $H_{0}>0$.  Instead, $\Xi$ reaches a maximum a finite distance from the horizon $r=r_{-}$ and then decreases to its Schwarzschild-de Sitter value at $r=r_{-}$.  As $H_{0}$ approaches zero, this maximum increases and at the same time its distance from the horizon decreases.  In this manner the curvatures on the actual horizon $r=r_{-}$ are bounded (and small if $m$ is large) for any $H_{0}>0$, but nevertheless the horizon becomes a null singularity at $H_{0}=0$.\footnote{This is reminscent of the Gibbs phenomenon encountered when one studies discontinuous limits of Fourier series representations.}  The order of limits where the cosmological constant is held fixed and one goes to the horizon at $t=\infty$, and {\it then} sends $H_0 \rightarrow 0$, regulates the singularity by first ``inflating away'' the FRW matter and pressure to zero due to the cosmological expansion.\footnote{Old age often leads to hair loss.}     No such mechanism is in operation when $H_0 = 0$ from the start.  Nevertheless as we have seen explicitly, the singularity that is present in the case of $H_0 = 0$ is very soft: first of all it is null, and secondly it is weak, being relegated to invariants which involve at least two derivatives of the Riemann tensor. Strictly speaking it does {\em not} violate the spirit of the strong cosmic censorship conjecture (see e.g. \cite{robm} for a nice review).

\section{Summary}
\label{sec:conclusions}

In this paper we analyzed the McVittie solutions of Einstein's field equations, which have been debated for nearly 80 years as candidates for describing the gravitational fields of spherically symmetric mass distributions in expanding FRW universes. For simplicity 
we focused on spatially flat McVittie geometries. Our results show that the McVittie solutions that asymptote to FRW universes dominated by a positive cosmological constant at late times are black holes with regular event horizons. While we were not able to construct the explicit form of the extension of the geometry past the black hole event horizon $r=r_-, t = \infty$, 
we showed that such analytic extensions {\em exist} when $H \rightarrow H_0 > 0$, being defined by the ingoing null geodesic family. Further, we demonstrated that in this limit the near horizon geometry is asymptotically Schwarzschild-de Sitter, which is essentially a consequence of the cosmic and black hole no-hair dynamics. Thus at least in this case, the shaded area past the surface $r=r_-, t = \infty$ in Figure \ref{sdsmcvitt} is a real black hole interior of some sort.

 When the exterior geometry asymptotes to an FRW cosmology with a power law scale factor, so that 
 $H = p/t \rightarrow 0$ in the future, the McVittie solutions are singular on the null surface $r=r_- = 2m, t = \infty$. This surface can still be reached by null geodesics  in finite affine parameter, and so is a real singularity. However it is very soft since it takes invariants involving at least two derivatives of the curvature to detect it. While it is naked, it respects the spirit of the strong cosmic censorship \cite{robm} in that it cannot be observed directly by external observers.
Our results correct the claims made in \cite{nolan} which have propagated through the 
literature.\footnote{See for example \cite{ers}. There are many conflicting incorrect claims, ranging from assertions that McVittie solutions are geodesically complete, over the suggestions that they cannot be black holes because of exotic spacelike singularities, to the  statement that McVittie solutions in any accelerating universes are black holes.  Our results establish a systematic method for determining what the McVittie solutions {\em are} and dispel those incorrect claims.}

Our most interesting conclusion is that the McVittie solutions in spacetimes which are asymptotically dominated by a positive cosmological constant are black holes. It seems rather remarkable that these black holes were not identified until nearly 80 years after the metric was written down. 
One might be puzzled that what seems to be an infrared quantity---the cosmological constant---regulates a short-distance property of the geometry, the would-be null singularity. This can be understood as a consequence of the special and restrictive form of the McVittie solutions, which requires the validity of a continuous fluid description down to the shortest scales, prevents the hole from accreting, and enforces that its energy density be exactly homogeneous on some set of spatial slices.   As we have discussed, these conditions are not particularly well-motivated physically. One could give up spatial homogeneity of the energy density, as in {\it e.g.} \cite{sussman} and its generalizations, or allow the hole to accrete. 
 
Another strategy for describing black holes in cosmological backgrounds is to go beyond the simple fluid description of cosmological matter and replace it with a microscopic description.
Such ideas motivated the work of \cite{kastra}, and were behind  the proposal of the levitating dark matter in \cite{kalpad}. A concrete example which realizes this has been found very recently in \cite{gibma}, based on the papers \cite{mae1,mae2}, and generalized in \cite{mae3}. The solutions found in those papers are very interesting, in that they encode large distance properties of the McVittie family. 
The matter content is a scalar field with an asymptotic runaway potential and a system of gauge fields with nonzero charges. Far from the hole the scalar rolling down the potential provides the source that drives the cosmic expansion. Near the hole the charge contributions correct the effective potential for the scalar and give it a large mass, employing essentially the same dynamics as the supersymmetric attractor mechanism in asymptotically flat black holes \cite{fixscalars}. Thus close to the black hole horizon the scalar decouples and there is no medium that could create divergences. This method of regulating the null McVittie singularity is local, but of course it requires abandoning both the fluid description near the black hole and the insistence of homogeneity of energy density outside of it, and is therefore completely consistent with our findings. It suggests that there might be other mechanisms for regulating the near horizon geometries of solutions which reduce to McVittie far away that might warrant further study.

\vskip.5cm

\smallskip

{\bf \noindent Acknowledgements}

\smallskip

We are grateful to B.~Carr, H.~C.~Cheng, A.~Gruzinov, M.~Luty, A.~Padilla, M.~Porrati and L.~Susskind 
for useful discussions, and to A.~Flachi for the help in obtaining a copy of the reference \cite{vaidya}.
We would also like to thank T.~Harada for questions on the previous version of the proof of finiteness of the $\tau$-time of Eq. (\ref{newtime}).
NK is grateful to RESCEU, University of Tokyo, and in particular to J. Yokoyama, for a kind hospitality during the course of this work. The work of NK was supported in part by the DOE
Grant DE-FG03-91ER40674.  The work of MK is supported by NSF CAREER grant  PHY-0645435.


\appendix

\section{The Short Happy Flight to $r = r_-, t = \infty$}
\label{app:infall}

Here we give the refined argument that the ingoing null geodesics reach the surface 
$r = r_-, t = \infty$ in finite affine parameter, which makes the McVittie geometry of (\ref{mcvitt}) geodesically incomplete to them. We further compute the precise rate at which the null radial geodesics approach the surface $r = r_-, t = \infty$, which we will need to show that in the limiting case with $H_0 \ne 0$ this surface is a regular event horizon. 

So to start with, note that the first of the Eqs. (\ref{radeq}) has a pole on the right-hand side there, which we can see by rewritting it as 
\be
\frac{dt}{dr} =  \frac{r \Bigl(\sqrt{1-2m/r} + Hr \Bigr)}{H^2 \sqrt{1-2m/r} (r-r_+) (r-r_-) (r+ r_+ + r_-)} \, ,
\label{poles}
\ee
where $r_\pm$ are the two positive roots of (\ref{metfunct}), and $- (r_+ + r_-)$ is its one negative root. 
In the limit $r \rightarrow r_-$, this yields
\be
dt \rightarrow - \frac{2r_- }{H_0^2 (r_+-r_-) (r_+ + 2r_-)} \frac{dr}{r-r_-(t)} \, ,
\label{polelimit}
\ee
where we in principle must take into account the fact that the roots $r_\pm$ of Eq. (\ref{metfunct}) are time-dependent, approaching the roots $r_\pm$ of (\ref{sdshor}) only as $t \rightarrow \infty$.

Equation (\ref{polelimit}) defines a differential equation that controls how the geometry approaches the surface $r = r_-, t =\infty$. Since we are interested in the limit $r \rightarrow r_-$, where the only divergence in (\ref{polelimit}) appears as the denominator of the RHS vanishes, while all other quantities remain finite, we can set $r_\pm$ to their asymptotic values everywhere except in the difference 
$r - r_-(t)$. This yields a first order linear differential equation
\be
\frac{dr}{dt} = - \frac{H_0}{\alpha} \Bigl(r - r_-(t)\Bigr) \, ,
\label{ode}
\ee
where from (\ref{polelimit}), we find $\alpha = \frac{2 r_-}{H_0 (r_+-r_-)(r_+ + 2r_-)}$.
Away from the extremal limit where $r_\pm$ coincide, and when the McVittie geometry asymptotes to the Schwarzschild-de Sitter limit with $H \rightarrow H_0 \ne 0$, integrating this is straightforward. To the leading order, when $m \ll H$ (with appropriate generalizations as $m$ approaches the extremal value), in the limit $t \rightarrow \infty$ we have $r_-(t) \rightarrow r_- + \Delta e^{-sH_0 t}$, where
$\Delta = 8 m^3 \rho (t=0)$. This follows from Eq. (\ref{friedman}) in the limit $t \rightarrow \infty$ and $m \ll H$, substituting for the energy density contents of the universe $\rho = \Lambda + \rho_0 \left( \frac{a_0}{a}\right)^s $, where $s = 3(1+w)$ and $w = p/\rho$ is the energy fluid equation of state, so that 
$H \rightarrow H_0 + {\cal O}(e^{-sH_0 t})$. Therefore the solution of (\ref{ode}) is
\be
(1-s\alpha) r_- e^{- H_0 t/\alpha} + \Delta e^{-sH_0 t} \rightarrow (1-s\alpha) \left({r-r_-} \right) + \ldots \, ,
\label{tofr}
\ee
to the leading order, where the ellipsis denote subleading terms,  and $\alpha >0$ reflecting the fact that $t \rightarrow \infty$ as $r \rightarrow r_-$. Clearly the rate at which $r$ approaches $r_-$ depends on the cosmological contents, since it is controlled by $1-\alpha s$. When $\alpha s < 1$ the evolution is dominated by the second term on the LHS, while when $\alpha s > 1$, the first term takes over. Further,
when $\alpha s =1$, the LHS of the solution (\ref{tofr}) degenerates into a resonance-like polynomial times the exponential $e^{-H_0 t}$, and while this case needs to be given special treatment, its generic behavior is the same as the nondegenerate cases. Moreover, one can check that in the extremal limit $r_+ = r_-$, which occurs when $m$ is very large, the right-hand side of  (\ref{eq:radial_geodesics}) has a double pole, implying that the divergence of $t$ is even faster, going as 
\be
H_0 t \rightarrow \frac{2}{3H_0 (r-r_-)} \, .
\label{fastdiv}
\ee
By the arguments above, $\dot H \propto e^{-sH_0t}$ in this limit.
Note, that $\dot H < 0$. Hence, along null ingoing radial geodesics near the null surface $r_-, t = \infty$, we have $\dot H \propto (r-r_-)$ when $ \alpha s  \le 1$ (with essentially irrelevant logs when $\alpha s = 1$, which we shall ignore) and $\dot H \propto (r-r_-)^{\alpha s}$ when $ \alpha s  >1$, and thus the coefficient of $r'^2$ on the right hand side of the second of Eqs. (\ref{radeq}) behaves as
\be
\frac{\dot H}{\sqrt{1-2m/r}(\sqrt{1-2m/r} - Hr)^2} \rightarrow - {\cal C}^2 (r-r_-)^p \, ,
\label{diverg}
\ee
where $p = -1$ when $\alpha s \le 1$ and $\alpha s - 2$ when $\alpha s >1$, and ${\cal C}^2$ is a positive constant given by the ratio of the energy fluid density and the cosmological constant. This expression is generically divergent for typical values of $\alpha s$. The asymptotic form (\ref{diverg}) enables us to integrate the second of Eqs (\ref{radeq}) along a null geodesic, to find that to the leading order
\be
r' = r_*' \exp\left(- {\cal C}^2 \int_{r_*}^{r} dr \left(\frac{r-r_-}{r_-}\right)^{p} \right) \, ,
\label{rprime}
\ee
where $r_*'$ is the initial radial `velocity' of an ingoing geodesics starting at some 
initial distance $r=r_* > r_-$. So we confirm that the geodesic `velocity' $r'$ is non-positive, and it might vanish when $r \rightarrow r_-$.
Integrating the exponent in (\ref{rprime}), we find that when $\alpha s \le 1$, so that $p=-1$,
\be
r' =  r'_* \Bigl(\frac{r_*-r_-}{r-r_-} \Bigr)^{{\cal C}^2 r_-} \, ,
\label{rprime11}
\ee
which clearly implies that $r'$ diverges to $-\infty$ as $r \rightarrow r_-$, since $r_*' < 0$. Therefore, there is no stopping of the fall towards $r_-$. Indeed, since the affine parameter is 
$\lambda = \int^r_{r_-} \frac{dr}{r'}$, we find that
\be
\Delta \lambda = \frac{1}{r'_*} \int^r_{r_-} dr \Bigl(\frac{r-r_-}{r_*-r_-} \Bigr)^{{\cal C}^2 r_-}  \, ,
\label{affineparameter1}
\ee
and since the integral is convergent, it takes a finite affine time $\Delta \lambda$ to get to $r=r_-$.
Similarly, when $\alpha s > 1$, so that $p=\alpha s - 2$, we have
\be
r' = {\cal A} \exp\left(- \frac{{\cal C}^2}{\alpha s-1} \left(\frac{r-r_-}{r_-}\right)^{\alpha s -1} \right) \, ,
\label{rprime1}
\ee
where ${\cal A} = r_*' \exp\left(\frac{{\cal C}^2}{\alpha s -1} \left(\frac{r_*-r_-}{r_-}\right)^{\alpha s -1} \right)$. So ${\cal A}$ is finite and strictly negative, since $r_*' < 0$. Therefore, since $\lambda = \int^r_{r_-} \frac{dr}{r'}$, we find that
\be
{\cal A} \Delta \lambda = \int^r_{r_-} dr \exp\left(\frac{{\cal C}^2}{\alpha s -1} \left(\frac{r-r_-}{r_-}\right)^{\alpha s -1} \right) \, .
\label{affineparameter}
\ee
Since the exponent vanishes as $r \rightarrow r_-$, the integral still converges, and again it takes a finite affine time to get to $r=r_-$. Hence, in any case the right-hand side of either (\ref{affineparameter1}) or (\ref{affineparameter}) is always finite, confirming our assertion that the average geodesic `velocity' $r'$ is negative definite. A similar proof exists when the asymptotic value of $H$ vanishes. This detailed analysis confirms our earlier argument, showing that ingoing radial null geodesics reach the null surface $r=r_-, t = \infty$ in a finite lapse of the affine parameter, and hence that the portion of the McVittie geometry (\ref{mcvitt}) bounded by this null surface, the singularity $\mu = 1$ and the exterior cosmological regime is {\it not} geodesically complete.

\section{The Curvature and Einstein Tensors}
\label{app:curvature}

It is straightforward to compute the full Riemann tensor for the McVittie solution (\ref{mcvitt}). Specializing to the coordinates $r, t$, and starting with the metric (\ref{cosmetric}), after a straightforward albeit tedious calculation we find that the nonvanishing components are
\begin{eqnarray}
R_{trtr} &=& - \frac{\dot H}{\sqrt{1-2m/r}} - \frac{2m}{r^3} - H^2 \, , \nonumber \\
R_{t\theta t \theta} &=& - \dot H r^2 \sqrt{1-2m/r} + \frac{m}{r} \Bigl(1-\frac{2m}{r}\Bigr) - H^2 r^2 (1 - \frac{m}{r} - H^2 r^2)  \, ,  ~~~~~
R_{t\phi r \phi} = R_{t\theta r \theta} \sin^2 \theta \, , \nonumber \\ 
R_{r\theta r \theta} &=& \frac{H^2 r^2 - m/r}{{1-2m/r}} \, ,   ~~~~~
R_{r\phi r \phi} = R_{r\theta r \theta} \sin^2 \theta \, ,  \\ 
R_{\theta \phi \theta \phi} &=& \Bigl(\frac{2m}{r} + H^2 r^2 \Bigr) r^2 \sin^2 \theta \, , ~~~~~
R_{t\theta r \theta} = - \frac{H r(H^2 r^2 - m/r)}{\sqrt{1-2m/r}} \, , ~~~~~
R_{t\phi r \phi} = R_{t\theta r \theta} \sin^2 \theta \, , \nonumber
\end{eqnarray}
and it is quite clear upon direct inspection that they are {\it all} well behaved in the limit $r = r_-, t = \infty$ as long as $H \rightarrow H_0 \ne 0$. 

The Einstein tensor is found from the contractions of the Riemann tensor. It is
\begin{eqnarray}
G^{{t}}{}_{{t}} &=& -3H^2 \, , \nonumber \\
G^{{r}}{}_{{r}} &=& G^{{\theta}}{}_{{\theta}} = G^{{\varphi}}{}_{{\varphi}} = -3H^2 - \frac{2}{\sqrt{1-2m/r}}\dot{H} \, .
\end{eqnarray}

\section{The Singular Invariant in the Power-law McVittie}
\label{app:xi}

Here we quote the explicit form of the invariant $\Xi \equiv (\nabla_{\mu} \nabla_{\nu} R_{\rho \sigma \delta \lambda})^{2} $:
\ba
\Xi &=& \Biggl[ 12 \sqrt{1-\frac{2 m}{r}} r^6  {H} ^4 \Bigl(-1320 m^2 (2 m-r)^5+r^7 (885 m^4-1686 m^3 r+1240 m^2 r^2-412 m r^3  \nonumber \\
&& +52 r^4 )    {\dot H}^2 \Bigr)-24 r^{13} \left(-158 m^4+185 m^3 r-63 m^2 r^2+m r^3+2 r^4\right)  {H}^3   {\dot H}{\ddot  H} \nonumber\\ && -24 (2 m-r) r^{10}  {H}    {\ddot H}  \biggl( m^2 \left(6 m^2-19 m r+8 r^2\right)   {\dot H}  +\sqrt{1-\frac{2 m}{r}} r^4  \left(67 m^2-56 m r+12 r^2\right)   {\dot H} ^2 \nonumber \\
&&+  r^4 \left(14 m^2-11 m r+2 r^2\right)  {\partial_{t}^{3} H} \biggr)   + 4 r^3 (-2 m+r)  {H} ^2 \biggl[-6 m r^{10} \left(57 m^2-47 m r+10 r^2\right)  {\dot H} ^3\nonumber  \\  
 &&
-2 m^2 \sqrt{1-\frac{2 m}{r}} r^7 \left(334 m^2-347 m r+96 r^2\right)    {\dot H} ^2
 +3 \sqrt{1-\frac{2 m}{r}} \Bigl(240 m^2 (25 m-12 r) (-2 m+r)^4 \nonumber  \\
 && +r^{11} \left(109 m^2-88 m r+
 19 r^2\right)   {\ddot H} ^2\Bigr)
  -6 m r^3   {\dot H}  \Bigl(180 m (-2 m+r)^4+\sqrt{1-\frac{2 m}{r}} r^8 (-5 m+r)  { \partial_{t}^{3} H} \Bigr)\biggr] \nonumber
 \\  
 && +4 (-2 m+r) \biggl[ 2 m^2 r^{10} \left(6 m^2+m r-2 r^2\right)   {\dot H} ^3+3 \sqrt{1-\frac{2 m}{r}} r^{14} \left(57 m^2-52 m r+12 r^2\right)   {\dot H} ^4 \nonumber
 \\&&
 +r^7 (-2 m+r)   {\dot H} ^2 \left(m^2 \sqrt{1-\frac{2 m}{r}} \left(847 m^2-832 m r+222 r^2\right)-6 (5 m-2 r) r^7  {\partial_{t}^{3}  H} \right)
\nonumber  \\ &&
 -2 m^2 (2 m-r) r^3   {\dot H}  \left(180 m (2 m-r)^3+7 \sqrt{1-\frac{2 m}{r}} r^8  {\partial_{t}^{3}  H} \right)
\nonumber \\&&
 +\sqrt{1-\frac{2 m}{r}} (-2 m+r) \Bigl(-40 m^2 r^{11}   {\ddot H} ^2+3 \bigl(-120 m^2 (2 m-r)^3 \left(65 m^2-60 m r+14 r^2\right) \nonumber \\
 && +r^{15}  {\partial_{t}^{3}  H} ^2\bigr)\Bigr)\biggr] \Biggr] (r^{17} (1-2 m/r)^{11/2} )^{-1} \, .
 \nonumber
\ea

\end{document}